\documentclass{aastex63}
\usepackage{graphicx}
\usepackage{amsmath}
\usepackage{lineno}
\usepackage{enumitem}

\def\Msolar{\ifmmode {\rm M_{\odot}}\else $\rm M_{\odot}$\fi}
\def\Mearth{\ifmmode {\rm M_{\oplus}}\else $\rm M_{\oplus}$\fi}
\def\Rearth{\ifmmode {\rm R_{\oplus}}\else $\rm R_{\oplus}$\fi}
\def\micron{\ifmmode {\mu{\rm{m}}}\else $\mu$m\fi}

\def\inc{\imath}

\begin{document}

\title{Light Echoes of Protoplanetary Disks}

\author[0000-0001-9181-1105]{Austin J. King}
\author[0000-0001-7558-343X]{Benjamin C. Bromley}
\affil{Department of Physics \& Astronomy, University of Utah, 
\\ 115 S 1400 E, Rm 201, Salt Lake City, UT 84112}
\email{austin.king@utah.edu}

\begin{abstract} 
Light echoes offer a means of studying protoplanetary disks, including their geometry and composition, even when they are not spatially resolved. We present a test of this approach applied specifically to optically thick, geometrically flared disks around active stars. Here we adopt stellar parameters of an active M dwarf to calculate light echoes for disks and rings with radii that would produce time-delays consistent with TESS short cadence (about 2 minute) time bins. Our results show successful fits to disk parameters, highlighting the potential effectiveness of this method in the search for protoplanetary disks.
\end{abstract}


\section{Introduction} 
\label{intro}
Active stars create an opportunity to observe surrounding stellar material that may otherwise be undetectable. Strong flares can brighten a planetary disk surrounding an active star \citep{gaidos1994}. Multiple stellar flares can produce cumulative light curves with sufficient signal-to-noise for the detection of light echoes --- scattered post-flare light from the disk \citep{Mann,bromley2021,King24}. This time-domain flux data can be compared against model disks to estimate disk geometry and orientation, as well as disk mass in the case of optically thin debris disks.

This vein of work is not new. Light echoes from impulsive flares have been seen as a potential method for planet detection \citep{argyle1974, bromley1992} and the study of circumstellar disks \citep{gaidos1994} for decades. Photo-reverberation mapping, which essentially detects the response of scattered light to intrinsic variations in stellar brightness \citep{Huan16}, has been applied to a broad range of systems, including protostars \citep[e.g.,][]{francis22} and active galactic nuclei \citep[e.g.,][]{shen2024}. Both photo-reverberation and light echoes refer to making observations of a star with variable brightness which leads to brightened surrounding material at some time-delay. The time-delay is a result of the distance from the star to the circumstellar material. Complex structural details of disks can also be determined from these methods; \cite{kama16} showed that spiral structures in protoplanetary disks can be determined from photon travel time delays.

In this paper, we revisit the disk structure used in \citet{gaidos1994}, presenting optically thick, geometrically flared protoplanetary disks, as opposed to \citet{bromley2021, King24} who modeled optically and geometrically thin debris disks. Protoplanetary disks are created in the early life of a stellar system. Following the birth of a low-mass star, the remaining matter of the collapsing molecular cloud forms an extended disk of cool dust and gas which persist for millions of years \citep{Williams11}. The nature of these disks makes them of particular interest for gaining information on young stellar systems and for our understanding of planet formation.

A host of studies \citep{Hartigan95, Strom89, Hartmann01, Haish01, hillenbrand05} regarding the lifespan of disks suggest that the hosts of protoplanetary disks are very young, and that the disks themselves are short-lived.
The majority (60-80\%) of stars younger than 1~Myr show evidence of protoplanetary disks, while those older than 10~Myr present this evidence in only about 10\% of cases. Recent exciting works focused on young systems with protoplanetary disks include a JWST proposal aiming to reveal physical and chemical changes disks experience due to variable accretion and high-energy irradiation \citep{Kospal24jwst}, the discovery of one of the largest known protoplanetary disk \citep{Ciprian24}, and exploration of the impact that energetic flares from T Tauri stars have on their surrounding protoplanetary disks, specifically how these flares contribute to the ionization of the inner regions of the disks \citep{Brunn24}. \cite{Huan16} worked with protoplanetary disks in a similar endeavor that we approach in this paper --- they utilized photo-reverberation to map the structure of a protoplanetary disk, hosted by a T Tauri star, and showed results for the first light echo detected from the inner wall of such a disk.

In this paper, we continue to build upon these works on protoplanetary disks by modeling light echoes produced from summed flare events and fitting these against model disks with variable geometric parameters and albedo. We find promising results on the extraction of disk parameters from artificial light echo data. Also, as discussed in \cite{King24}, this method does not require disks to be resolved from their host star, thus opening a new door for detection of protoplanetary disks in time-domain data.

\section{Modeling Light Echoes}

\subsection{\cite{King24} Summary}\label{k24sum}
The approach to forming model echoes follows closely with \cite{King24}. The following is a summary of how time-series light curves were produced for optically thin debris disks in this paper.

The Draine phase function \citep{draine2003} is utilized to determine angle-dependent intensity of light scattered off of astrophysical gas or dust. In equation form, this function is defined as
\begin{equation}\label{eq:draine}
    \Phi_\alpha(\theta_s) = 
    \frac{1}{4\pi} 
    \left[\frac{(1-g^2)}{(1+\alpha(1+2g^2)/3}\right] \frac{1+\alpha\cos^2\theta_s}{(1+g^2-2g\cos{\theta_s})^{3/2}},
\end{equation} 
where the parameters are set as $g=0.429$ and $\alpha=0.114$, in accordance with \cite{draine2003} results for 0.8 $\mu m$ light, interacting with forward scattering dust.

The scattering angle, $\theta_s$, is calculated by 
\begin{equation}
    \theta_s = \arccos{\left(-\sin{\phi}\sin{i}\right)}
\end{equation}
where $phi$ is the angular position on the disk, measured from $0-2\pi$ (0 is marked on the positive x-axis projected on the sky) and $i$ is the disk inclination, measured from $0^\circ-90^\circ$ ($0^\circ$ for a face-on disk, $90^\circ$ for edge-on).

We can now define the brightness of any angular bin ($dA=rdrd\phi$) as 
\begin{equation}\label{eq:B}
    B = \frac{C(r) dA}{r^2}\left|1-\left|\phi/\pi-3/2\right|\right|\Phi_\alpha(\theta_s)
\end{equation}
The absolute value term is a result of \textbf{angle-averaging flare locations over multiple events}; flares on the near-side of the star are more likely to be observed, so preference is given to this position.
Note: this expression corrects a typographical error in \cite{King24} - with the defined zero point of $\phi$ on the positive x-axis, we re-write $\left|1-\phi/\pi\right|$ as $\left|1-\left|\phi/\pi-3/2\right|\right|$. This results in a maximum value on the near side of the disk ($\phi=3\pi/2$) and minimum on the far side of the disk ($\phi = \pi/2$).

In Equation \ref{eq:B}, the dependence of the scattered light on several key physical parameters of the system is contained within the value $C$, which is defined as
\begin{equation}
    C(r) =\frac{3 Q_\text{eff} M_\text{disk}}{8\pi\rho X r_\text{p,min}(r_\text{out}-r_\text{in})r},
\end{equation}
In the above equation, $Q_\text{eff}$ is the scattering efficiency of individual dust grains (taken as $Q_\text{eff}=2$, consistent with scattering theory for micron-size or larger grains \citep{vandehulst}), the parameter $X$ is included due to use of average dust properties instead of the underlying size distribution ($n(r_p)$). Minimum and maximum particle sizes of $r_{p,min} = 1$~$\mu$m and $r_{p,max} = 1$~mm, were adopted so that $X \approx 30$, where $X$ is the size distribution of the dust particles, and is defined as $X=(\frac{r_{p,max}}{r_{p,min}})^{1/2}$ \citep[cf.]{bromley2021}.

\eqref{eq:B} can then be paired with the crossing time, defined by \citet{gaidos1994}:
\begin{equation}\label{eq:T}
    T=\frac{r}{c}\left(1+\sin{\phi}\sin{i}\right).
\end{equation}
The disk is divided radially and azimuthally into small patches (area $dA$) and each patch is assigned a characteristic light travel time, $T$, from the above equation. Brightness from all patches with $T$ falling within some time bin centered at $t_j$ are combined to get the total flux of reflected light in that time bin. This process is repeated for all patches across the surface of the disk to generate a complete light curve. The parameters $M_\text{disk}$, $r_in$, $r_out$, and $i$ are left as the defining physical parameters of the debris disk light curve.

\subsection{Light Echo Calculation}

This paper follows closely with the above summary, with two significant changes. First, we assume isotropic scattering off of an optically thick disk \citep{gaidos1994}, which simplifies Equation \ref{eq:draine} to $\frac{1}{4\pi}$. Second, due to the nature of optically thick disks, it is not possible to probe their interiors as could be done with optically thin debris disks. Therefore, we drop the $C$ term from Equation \ref{eq:B}, rendering us unable to include mass as a fit parameter. We instead replace this with albedo. The remaining parameters are geometric properties of inner radius, outer radius, and inclination ($r_{in}, r_{out},$ and $i$). For our formalism, we maintain that $i=0^\circ$ indicates a face-on orientation, and $i=90^\circ$ indicates edge-on.

In addition to the above changes, we model a flared disk with a physical height, $H$ dependent on radius, as opposed to a geometrically thin disk. The extent of the disk follows a scale height relationship of $h=H/r_M$, where $h=0.045$ for disks with extent under 10 AU \citep{Bitsch15}. 

\begin{figure}[!h]
    \centering
    \includegraphics[scale=0.7]{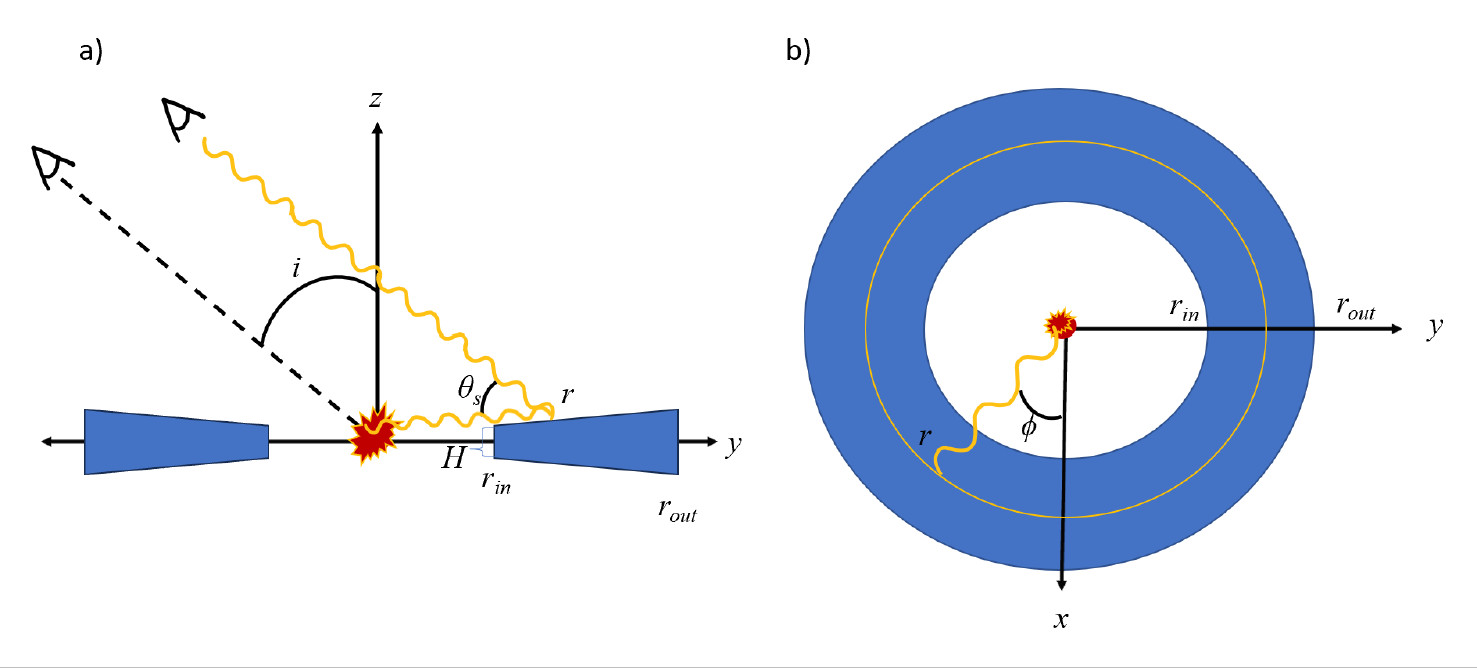}
    \caption{Geometric Disk Model: a) A cross-sectional side view along the x-axis of the protoplanetary disk. The flash at the origin depicts a solar flare. The yellow squiggles show the path of a photon as it strikes at a distance $r$ on the disk and reflects toward the observer at a scattering angle of $\theta_s$. $i$ denotes the inclination of the disk relative to the observer ($i=0^\circ=$ face-on, $i=90^\circ=$ edge-on), $H$ is the physical height of the disk at its inner radius, $r_{in}$ and $r_{out}$ is the outer radius. b) a top-down view of the disk along the z-axis. $r_{in}$ and $r_{out}$ once again denote inner and outer radii, respectively. The yellow ring shows all angles, $phi$, that a photon traveling to distance $r$ will interact with before scattering to the observer. Note: In this orientation, the portion of the disk aligned with the negative y-axis would be defined as the "near-side", as it is inclined toward the observer.}
    \label{fig:geometry}.
\end{figure}

Figure \ref{fig:geometry} shows the geometry of the system, depicting the flared structure of the disk and important variables. With this in mind, we can begin to describe the equations used to calculate model echoes.

With these changes in mind, we now redefine the equations listed in Section \ref{k24sum} as follows:

The Draine phase function \citep{draine2003} in its entirety is shown in Equation \ref{eq:draine} For this work, we set $g=\alpha=0$, which reduces the above equation to $1/{4\pi}$, consistent with isotropic scattering of reprocessed light.

The relative brightness of each angular bin with $dA=rdrd\phi$ is then defined as
\begin{equation}\label{eq:B_new}
    B = \gamma\frac{dA}{4\pi r^2}\left|1-\sin(i)*\left|\phi/\pi-3/2\right|\right|
\end{equation}
where $\gamma$ denotes albedo, replacing $C$ in Equation \ref{eq:B}, and $\phi$ and $r$ represent the angular and radial positions on the disk, respectively. $r$ is calculated from the scale height relationship. $r = \sqrt{r_M^2 + H^2}$, where $H=r_M\times h$. We also include a new dependence on inclination involved in determining the angle-average position of flares. We include the scaling term $\sin(i)$, such that in a face-on orientation, there is no $\phi$ dependence - all angular positions at the same $r$ should receive the same average illumination.

Finally, we utilize the same crossing time described in Equation \ref{eq:T}. We can then take the brightness of each physical bin from Eq. \ref{eq:B_new} and sum those with a shared time bin (Eq. \ref{eq:T}). This results in a Flux-Time lightcurve produced from any desired disk parameters. As in \cite{King24}, we envision coadded post-flare light curves as a means of increasing signal-to-noise. This leads to light curves comparable to those produced by isotropic flares, removing sensitivity to the location of flares on the stellar surface. 

To make our light curves further replicate real data, we look to TESS sources as a guide. Utilizing standard TESS errors as our $1\sigma$ uncertainty and the Python routine \texttt{numpy.random.normal}, we approximate noise and add it to each flux bin. Note that for ring-like structures, we required 1/3 standard TESS errors to produce viable data. More powerful instruments like the James Webb Space Telescope are capable of recording data at this lower-error level \citep{carter24}.

\begin{figure}[!ht]
    \centering
    \includegraphics[width=.44\textwidth, height=.21\textwidth]{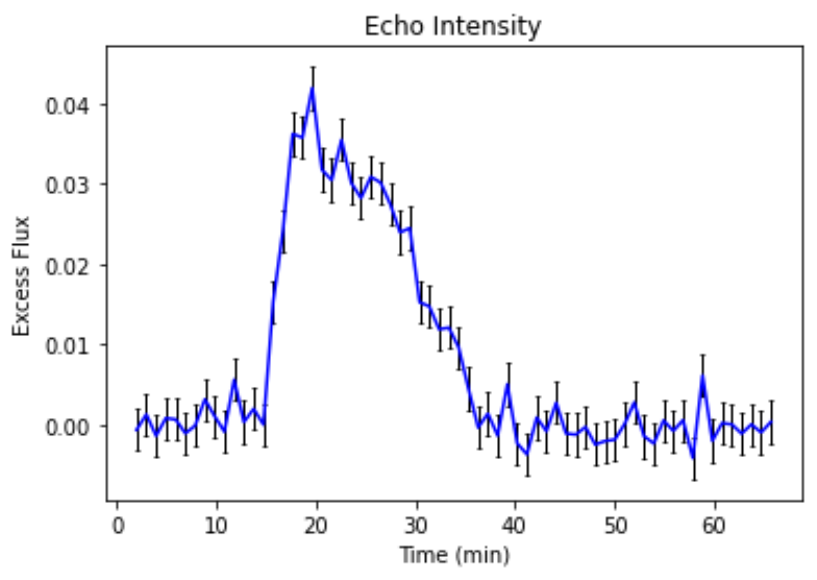}
    \includegraphics[width=.44\textwidth, height=.21\textwidth]{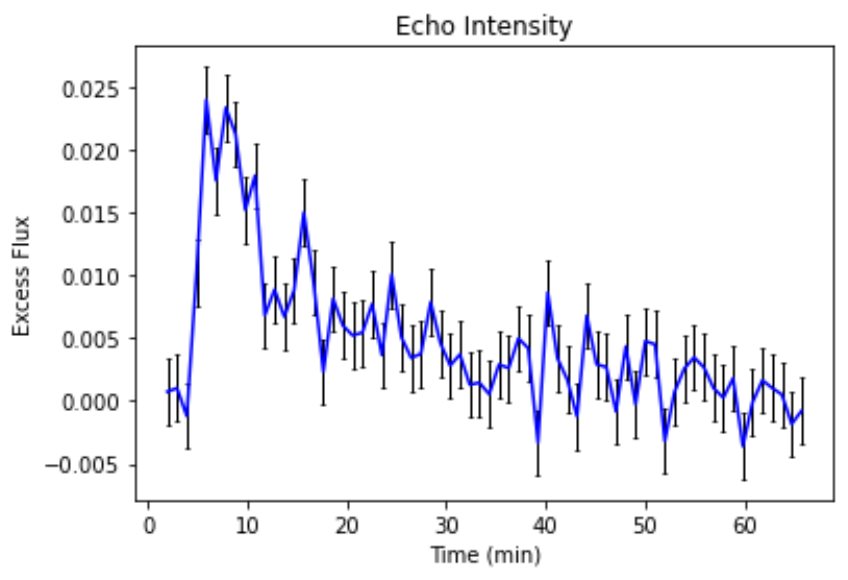}
    \includegraphics[width=.44\textwidth, height=.21\textwidth]{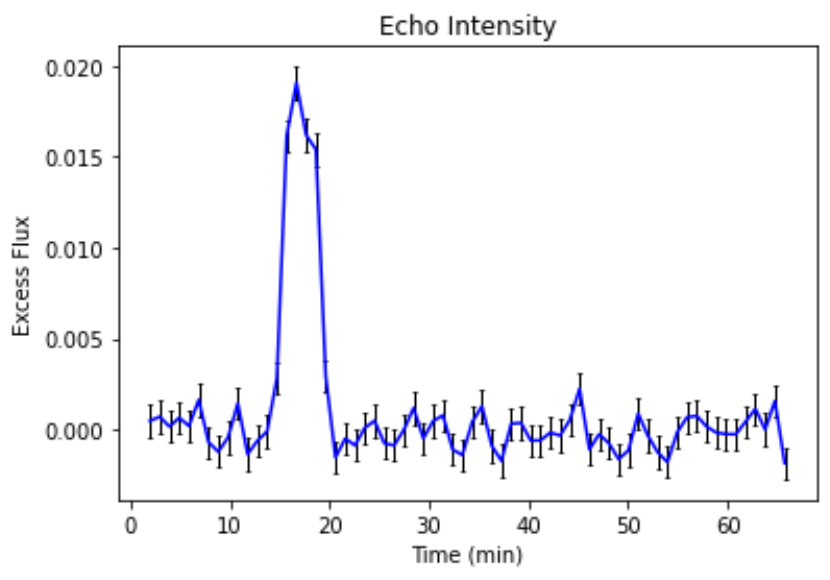}
    \includegraphics[width=.44\textwidth, height=.21\textwidth]{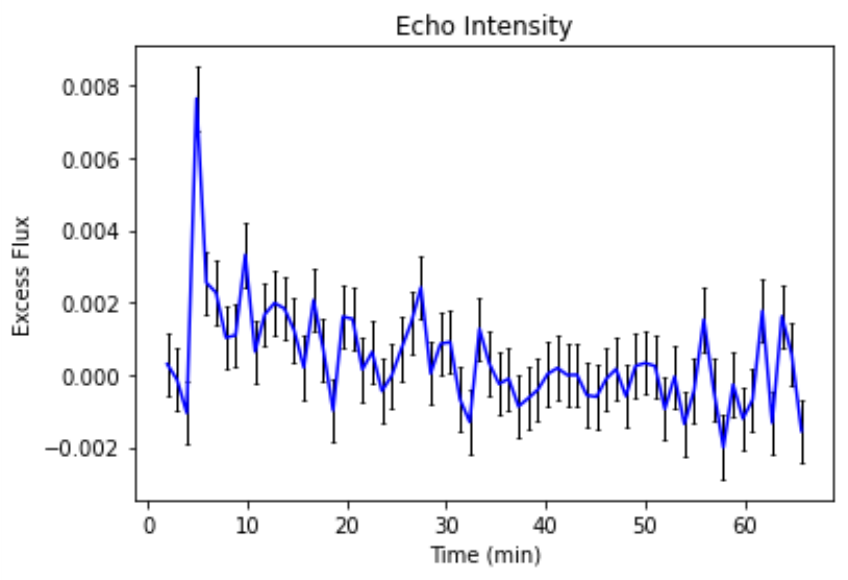}
    \caption{Model Light Echoes: Each of the panels above show the shape of a modeled echo light curve from an optically thick, circumstellar structure. The top two panels show a more extended disk, modeled with inner radius $r_{in}=2$ AU and outer radius $r_{out}=2 r_{in}$. The bottom panels show the same information for a ring-like structure, with $r_{in}=2$ AU and $r_{out}=1.1 r_{in}$. The two left panels represent a near face-on disk ($i=5^\circ$) and the two right panels represent a disk inclination of $i=45^\circ$. All curves are modeled with an albedo of 0.5.}
    \label{fig:curves}
\end{figure}

We are now able to produce realistic light curve data similar to the short-cadence exposures taken by TESS. This allows us to probe the inner regions of a stellar system, at distances less than about 10 AU. Model light curves for 4 different geometric orientations are shown in Figure \ref{fig:curves}. These echoes are formed from model disks with near face-on inclinations (top two panels), $i=45^\circ$ inclinations (bottom two panels). The left two panels depict a thinner ring-type geometry ($r_{out}/r_{in}=1.1$) and a slightly larger disk-type geometry ($r_{out}/r_{in}=2.0$. We chose to set $r_in = 2 \text{AU}$, landing our disk and ring sizes within the smaller end of disk sizes derived by \cite{Trapman_2023}. We keep the inner radius somewhat distant from the host star as this aids us in achieving a significant enough time delay between the arrival of flare light and the earliest arriving echo light, so that the two may be distinguished.

The echoes shown in the disk-like geometry (top two panels) have excess flux peaks of $f\approx0.04$ (near face-on) and $f\approx0.025$ ($i=45^\circ$), and their total increases in flux over the quiescent background, integrated over the time of the echo, are approximately $50\%$ and $30\%$, respectively. The peaks for ring-like geometry (bottom two panels) are found to be $f\approx0.02$ and $f\approx0.008$, and produce total flux increases of about $8\%$ and $5\%$.

\section{Fitting Light Echoes}

With the structure in place to create realistic light curves, we now turn to fitting specific disk parameters as was done in \cite{King24}. For this work, we chose disk parameters of $r_{in}=2$ AU, $r_{out}=1.1 r_{in}$, and $i=0^\circ, 45^\circ$ for ring-like structure and $r_{in}=2$ AU, $r_{out}=2.0 r_{in}$, and $i=0^\circ, 45^\circ$ for disk-like structure. Curves for both structures are calculated with an albedo of $0.5$. Using these values to create artificial data of light curves, we pass them through a $\chi^2$ goodness of fit test against various model disks:
\begin{equation}\label{eq:chi2}
    \chi^2 = \sum_{j=1}^{N}\left(F_{data, j} - F_{model}(t_j-t_0, r_{in}, r_{out}, i)\right)^2/\sigma_j^2
\end{equation}
where $j$ denotes the time bin index and $t_0$ is the time of flare occurrence. This process takes the input data light curve and compares it against curves generated randomly in parameter space to determine best-fit parameters.

We then assume independent Gaussian errors and define our log probability as 
\begin{equation}\label{eq:logP}
    \ln{\mathcal{P}} = -\frac{1}{2}\chi^2 -\frac{1}{2}\sum_{i}\ln\left({2\pi\sigma_i^2}\right) + \ln{\mathcal{P}_{priors}}
\end{equation}
where $\mathcal{P}_{priors}$ is defined as
\begin{equation}
    {\cal P}_\text{priors} \sim  
    \begin{cases}
    \ \sin(\inc) \ & \ \ 0^\circ \leq \inc \leq 90^\circ \ \text{and} \\
    & r_{min} \leq r_{in} < r_{out} \leq r_{max}, 
    \\
    0 & \text{otherwise.}
    \end{cases} 
\end{equation}
The geometric factor $\sin(i)$ accounts for the likelihood of observing disks in specific orientations. A randomly selected disk is more likely to be observed in some approximately edge-on configuration rather than directly face-on. We also use an uninformative prior on albedo limiting it to values between zero and one.

\begin{figure}[!ht]
    \centering
    \includegraphics[width=.44\textwidth, height=.3\textwidth]{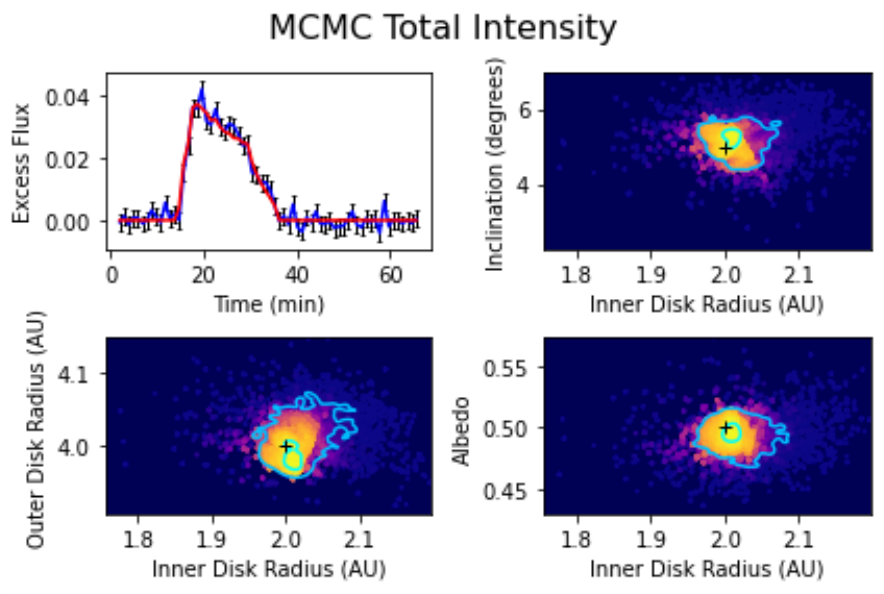}
    \includegraphics[width=.44\textwidth, height=.3\textwidth]{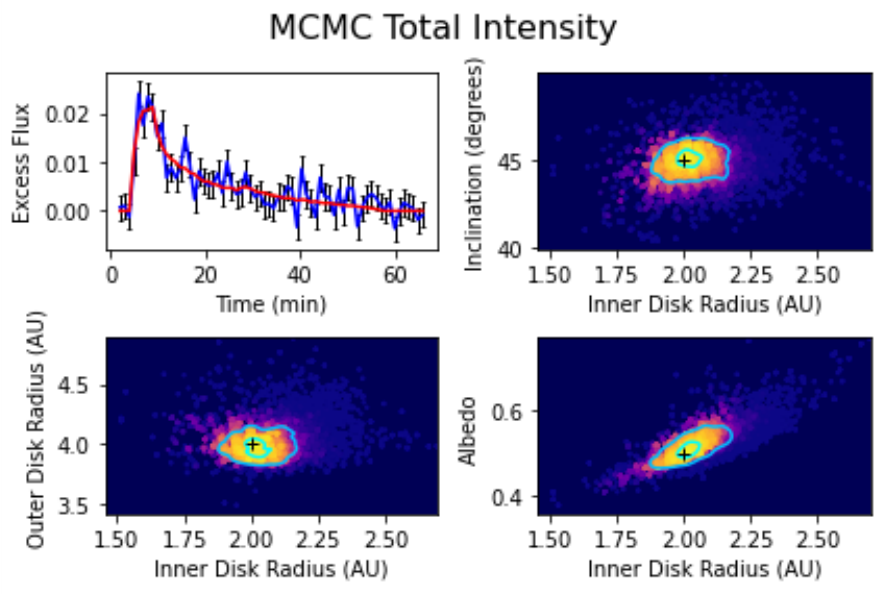}
    \includegraphics[width=.44\textwidth, height=.3\textwidth]{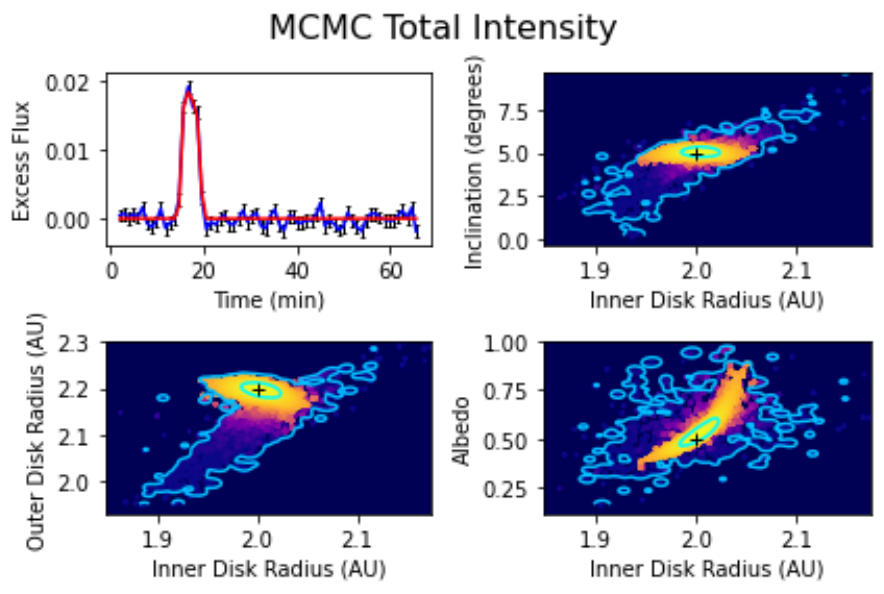}
    \includegraphics[width=.44\textwidth, height=.3\textwidth]{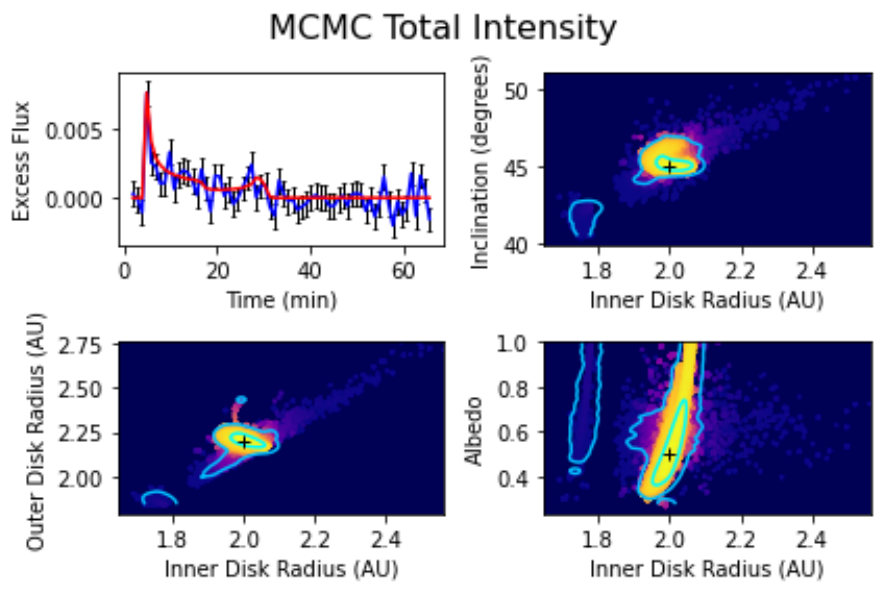}
    \caption{MCMC Echo Fits: Each of the panels above show the parameter space fit of the modeled echo light curve from Figure \ref{fig:curves}. The top two panels show fits for disk-like structure. The bottom panels show the same information for ring-like structure. The two left panels represent a near face-on disk ($i=5^\circ$) and the two right panels represent a disk inclination of $i=45^\circ$. We show confidence boundaries of $1\sigma$ (light blue) and $3\sigma$ (cyan) and true parameters are denoted with a black "+".}
    \label{fig:fits}
\end{figure}

We can then map out the likelihood of any given disk geometry in parameter space by utilizing these likelihood parameters in the Markov-Chain Monte Carlo package emcee \citep{emcee2012}. The burn-in stage is ran for the first 20 steps, after which the positions of the walkers are saved and the sampler is reset. We utilized 600 walkers which are given 4000 steps. This surpasses the auto-correlation time of our parameters. The initial ranges for the walkers are as follows: $r_{in}: 0-10 \text{ AU}$, $i: 0^\circ-90^\circ$, $\text{albedo}: 0-1$. The only constraint placed on $r_{out}$ is $r_{out} > r_{in}$. Figure \ref{fig:fits} shows the results of this process. The plots are organized by disk parameter in the same way as in Figure \ref{fig:curves}.

In each of these cases, we see very little deviation in the MCMC fitting, indicating high precision and correct extraction of disk parameters. With this particular model, we opted not to utilize edge-on orientations, as it would be particularly difficult to untangle echo light from the flare itself when there is little to no time delay. This is discussed in \cite{King24} for debris disks, where the use of polarization can improve this method's ability to separate flare light from echo light. 

\section{Summary and Discussion}
The utilization of post-flare light echoes opens the door for study of protoplanetary disks in the time domain. In work, we focused on the use of maximum likelihood fitting to return geometric parameters for optically thick disks and showed promising theoretical results, shown in Figure \ref{fig:fits} and summarized below:

\begin{enumerate}[topsep=2pt, partopsep=2pt, itemsep=10pt,parsep=2pt]
    \item Excellent fit results for disk-like structure:
    \subitem near face-on: uncertainty in $r_{in}$ of $\sim 0.2$ AU, $r_{out}$ of $\sim 0.2$ AU, $i$ of $\sim 2$ degrees and albedo of $\sim 0.1$.
    \subitem inclined: uncertainty in $r_{in}$ of $\sim 0.5$ AU, $r_{out}$ of $\sim 0.5$ AU, $i$ of $\sim 5$ degrees and albedo of $\sim 0.2$.
    
    \item Great fit results for ring-like structure geometric parameters:
    \subitem near face-on: uncertainty in $r_{in}$ of $\sim 0.2$ AU, $r_{out}$ of $\sim 0.3$ AU, and $i$ of $\sim 5$ degrees.
    \subitem inclined: uncertainty in $r_{in}$ of $\pm 0.4$ AU, $r_{out}$ of $\sim 0.5$ AU, and $i$ of $\sim 6$ degrees.

    \item Less constrained uncertainty for albedo in ring-like structures:
    \subitem near face-on: uncertainty of $\sim 0.7$
    \subitem inclined: uncertainty of $\sim 0.7$ 
\end{enumerate}

We conclude that both detection of unresolved disks and rings from post-flare data and determination of geometric parameters and albedo is possible. We plan to put this hypothesis to the test in additional works.

Future work may uncover compositional characteristics of protoplanetary disks. \cite{Sparks2018} showed how scattered light can reveal composition through spectral data, albedo, and polarization. These measurements, however, require precision that is not currently feasible. Future telescopes could provide the accuracy necessary to allow detection of composition.

Our next phase of research on this topic will focus on applying this method to data taken from flaring stars with known disks. This will provide a direct test of the light echo method's ability to return correct disk parameters from real data. 

\section{Acknowledgments}
We thank an anonymous referee for suggestions to improve this paper.

\pagebreak
\bibliography{bib}{}

\end{document}